\documentstyle[multicol,aps,prb]{revtex} 

\begin{document}
%\draft
\title{Shot-noise suppression in Schottky barrier diodes.}
\author{G. Gomila and L. Reggiani}
\address{Dipartimento di Ingegneria dell' Innovazione e Istituto Nazionale\\
di Fisica della Materia, Universit\'{a} di Lecce, Via Arnesano s/n, \\
73100 Lecce, Italy}
\author{J. M. Rub\'{\i}}
\address{Departament F\'{\i}sica Fonamental, Universitat de Barcelona, \\
Av.\ Diagonal 647,\\
08028 Barcelona, Spain}
\maketitle

\begin{abstract}
We give a theoretical interpretation of the noise properties of Schottky
barrier diodes based on the role played by the long range Coulomb
interaction. We show that at low bias Schottky diodes display shot noise
because the presence of the depletion layer makes negligible the effects of
the Coulomb interaction on the current fluctuations. When the device passes
from barrier to flat band conditions, the Coulomb interaction becomes
active, thus introducing correlation between different current fluctuations.
Therefore, the cross-over between shot and thermal noise represents the
suppression due to long range Coulomb interaction of the otherwise full
shot-noise. Similar ideas can be used to interpret the noise properties of
others semiconductor devices.
\end{abstract}


\begin{multicols}{2}
\narrowtext

%\vspace*{1.0truecm}

It is well known that, in the absence of $1/f$ contributions, excess noise
in Schottky barrier diodes (SBD) exhibits shot noise followed by thermal
noise (neglecting hot-carrier effects).\cite{Ziel86,Trippe86} Shot noise is
attributed to the presence of a barrier which controls the exponential
increase of current with applied voltage. By contrast, the thermal behavior
is attributed to the series resistance which controls the more or less
linear increase of the current at the highest voltages. A phenomenological
approach to describe the cross-over from shot to thermal behaviors is to
express the current spectral density, $S_{I}$, as the partition between two
noise generators as:\cite{Ziel86,Trippe86} 
\begin{equation}
S_{I}=\frac{2q\overline{I}R_{j}^{2}}{(R_{j}+R_{s})^{2}}+\frac{4k_{B}TR_{s}}{%
(R_{j}+R_{s})^{2}}\text{,} 
\end{equation}
where $\overline{I}$ is the average electric current, $q$ the electron
charge, $R_{j}$ the junction resistance, $R_{s}$ the series resistance, $%
k_{B}$ Boltzman's constants and $T$ the temperature.

Microscopic approaches at hydrodynamic\cite{Gomila98,Gomila99} and kinetic
levels\cite{Gonzalez97,Gruzinskis98} confirm the above expression without
resorting to two independent noise sources, but providing a self-consistent
solution of the dynamics together with the Poisson equation. However, a
physical interpretation of the noise properties of SBD, and in particular of
the cross-over between shot and thermal noise, is still lacking in the
current literature and constitutes an intriguing question.

The aim of this paper is to address this issue by showing that the
cross-over between shot and thermal noise behaviors represents the
suppression due to long range Coulomb interaction of the otherwise full
shot-noise. This suppression occurs in SBD when the device passes from
barrier to flat band conditions because of the following reason. By
increasing the free carrier concentration the originally negligible
screening effects due to the Coulomb interaction become relevant, thus
acting as a shot noise suppressor. Accordingly, when controlled by screening
effects carrier number fluctuations give a negligible contribution to the
total noise power, which is now practically given by the velocity
fluctuations only, thus providing thermal noise.

The starting point of our approach is the non-degenerate fluctuating
drift-diffusion equation which, for the case of one dimensional geometries
and low frequencies, reads,\cite{Vliet94} 
\begin{equation}
\frac{I(t)}{A}=qn(x,t)\mu E(x,t)+qD\frac{dn(x,t)}{dx}+\frac{\delta I_{x}(t)}{%
A}\text{,}  \label{DrifDif}
\end{equation}
where $I$ is the instantaneous electric current, $A$ the cross-sectional
area, $n$ the number density, $\mu $ the mobility, $E$ the electric field,
and $D$ the diffusion coefficient. Moreover, $\delta I_{x}(t)$ is a Langevin
diffusion noise source that accounts for current fluctuations, and has zero
mean and correlation function\cite{Vliet94} 
\begin{equation}
\overline{\delta I_{x}(t)\delta I_{x^{\prime }}(t^{\prime })}=2Aq^{2}D%
\overline{n}(x)\delta (x-x^{\prime })\delta (t-t^{\prime })\text{.}
\label{CorrD}
\end{equation}
Henceforth, a bar denotes an averaged quantity. In previous equations the
time variable just specifies that the quantities are changing with time due
to the existence of fluctuations. Equation (\ref{DrifDif}) has to be
completed with Poisson equation and with appropriate boundary conditions to
take into account the long range Coulomb interaction and the thermionic
emission processes, respectively. For the rectifying contact the boundary
condition reads\cite{Gomila98} 
\begin{equation}
n_{0}(t)-n^{eq}=\frac{I(t)-\delta I_{TE}(t)}{qAv_{r}}\text{,}  \label{n(t)}
\end{equation}
where $v_{r}$ is the recombination velocity of the contact, $%
n^{eq}=N_{C}e^{-q\beta \phi _{bn}}$ the contact equilibrium density, with $%
N_{C}$ being the effective density of states in the conduction band, $\phi
_{bn}$ the barrier height, and $\beta =1/k_{B}T$, and $\delta I_{TE}(t)$ is
an additional Langevin noise source that accounts for the fluctuations due
to the random emission of carriers. It has zero mean and spectral density%
\cite{Gomila98} 
\begin{equation}
S_{I}^{TE}=2q\left[ \overline{I}+2I_{C}\right] ,  \label{SITE}
\end{equation}
with $I_{C}=qn^{eq}v_{r}A$. For the ohmic contact, if the sample is long
enough we can neglect their effects and take 
\begin{equation}
n_{L}(t)=N_{D}\text{.}  \label{nL(t)}
\end{equation}

By dividing Eq.(\ref{DrifDif}) by the wronskian 
\begin{equation}
W(x,t)=e^{-q\beta \int_{0}^{x}E(x^{\prime },t)dx^{\prime }}\text{,}
\label{W}
\end{equation}
and integrating the result between $x=0$ and $x=L$, where $L$ is the sample
length, one obtains (see also Refs. \onlinecite{Gomila98,Ziel78} ) 
\begin{equation}
I(t)=I_{L}(t)-I_{0}(t)+\delta I_{D}(t)\text{,}  \label{I(t)}
\end{equation}
where 
\begin{equation}
I_{0}(t)=qAv_{D}(t)n_{0}(t)\text{;\quad }I_{L}(t)=qAv_{D}(t)\frac{n_{L}(t)}{%
W_{L}(t)}\text{,}  \label{IL0(t)}
\end{equation}
and 
\begin{equation}
\delta I_{D}(t)=\overline{v}_{D}\int_{0}^{L}\frac{\delta I_{x}(t)}{D%
\overline{W}(x)}dx\text{.}  \label{dID(t)}
\end{equation}
To arrive at the previous results we have made use of Einstein's relation $%
D=\mu k_{B}T/q$, and have introduced the diffusion velocity as 
\begin{equation}
\frac{1}{v_{D}(t)}=\int_{0}^{L}\frac{dx}{DW(x,t)}\text{.}  \label{vD(t)}
\end{equation}
From Eq.(\ref{I(t)}), the average current is given by 
\begin{equation}
\overline{I}=\overline{I}_{L}-\overline{I}_{0}=I_{s}\left( e^{q\beta 
\overline{V}}-1\right) \text{,}  \label{Iav}
\end{equation}
where $I_{s}=qn^{eq}\frac{v_{r}\overline{v}_{D}}{v_{r}+\overline{v}_{D}}A$
and $\overline{V}$ is the applied bias. In deriving Eq.(\ref{Iav}) we have
used the averaged version of Eqs. (\ref{n(t)}), (\ref{nL(t)}) and (\ref
{IL0(t)}), and the fact that $\overline{W}_{L}=\exp \left( -q\beta (%
\overline{V}-V_{bi})\right) $, with $V_{bi}=k_{B}T/q\ln (N_{D}/n^{eq})$
being the built-in voltage.

Notice that in the definition of $\overline{v}_{D}$ we have included all the
sample (and not only the depletion layer as usual). As a result, the formal
expression for the $I-V$ characteristics holds in the whole range of applied
bias. For low applied bias $I_{s}$ is nearly constant, and the $I-V$
characteristics is exponential. At larger applied bias (above flat band
conditions), $I_{s}$ depends on $\overline{V}$ in such a way as to cancel
the exponential behavior and give rise to a linear characteristics, as
should be.\cite{remark}

The current fluctuations can be obtained by linearizing Eq.(\ref{I(t)})
around the stationary state. After some algebra one arrives at 
\begin{equation}
\delta I(t)=\delta I_{L}^{*}(t)-\delta I_{0}^{*}(t)+\delta I_{E}(t)+\delta
I_{D}(t)\text{,}  \label{dI(t)}
\end{equation}
where 
\begin{equation}
\delta I_{0}^{*}(t)=qA\overline{v}_{D}\delta n_{0}(t)\text{;\quad }\delta
I_{L}^{*}(t)=qA\frac{\overline{v}_{D}}{\overline{W}_{L}}\delta n_{L}(t)\text{%
,}  \label{dIL0(t)}
\end{equation}
and

\begin{equation}
\delta I_{E}(t)=\int_{0}^{L}H(x)\delta E_{x}(t)\text{,}  \label{dIE(t)}
\end{equation}
with 
\begin{equation}
H(x)=-Aq^{2}\beta \overline{v}_{D}\frac{\overline{n}(x)}{\overline{W}(x)}%
=-q^{2}\beta A\overline{v}_{D}N_{C}e^{\beta (F_{n}(x)-E_{C}(0))}\text{.}
\label{H(x)}
\end{equation}
Here $E_{C}(0)$ is the edge of the conduction band and $F_{n}(x)$ the
quasi-Fermi level. By substituting the linearized version of Eqs. (\ref{n(t)}%
) and (\ref{nL(t)}) into Eq. (\ref{dIL0(t)}), and substituting the result
into Eq.(\ref{dI(t)}), one finally obtains 
\begin{eqnarray}
\delta I(t) &=&\frac{\overline{v}_{D}/v_{r}}{1+\overline{v}_{D}/v_{r}}\delta
I_{TE}(t)+\frac{1}{1+\overline{v}_{D}/v_{r}}\delta I_{D}(t)+  \nonumber \\
&&\frac{1}{1+\overline{v}_{D}/v_{r}}\delta I_{E}(t)\text{,}  \label{dISch(t)}
\end{eqnarray}
The meaning of the different contributions to the current fluctuations in
Eq.(\ref{dISch(t)}) is the following:

(i) The term involving $\delta I_{TE}(t)$ represents the contribution to the
current fluctuations of the thermionic emission processes.

(ii) The term involving $\delta I_{E}(t)$ describes the current fluctuations
resulting from a response to an electric field fluctuation. This term
accounts for the effects of the long range Coulomb interaction on the
current fluctuations. To obtain their statistical properties one needs to
compute explicitly the electric field fluctuations.\cite{Gomila98,Gomila99}
For the present paper it is enough to notice that $\delta I_{E}(t)$ is
negligible when $H(x)$ is constant (or nearly constant) , since then $\delta
I_{E}(t)\approx H\int_{0}^{L}dx\delta E_{x}(t)=H\delta V(t)=0$, where we
have used that current fluctuations are calculated under fixed bias
conditions.

(iii) The term involving $\delta I_{D}(t)$ represents the current
fluctuations due only to diffusion processes since it is directly related to
the fundamental diffusion noise source, $\delta I_{x}(t)$. From Eqs.(\ref
{CorrD}) and (\ref{dID(t)}) it can be shown that the current fluctuations
associated to $\delta I_{D}(t)$ always display a low frequency spectral
density, $S_{I}^{D}=2\int_{-\infty }^{+\infty }dt\overline{\delta
I_{D}(t)\delta I_{D}(0)}$ , of the shot noise type (see Refs. %
\onlinecite{Gomila98,Ziel78} ) 
\begin{eqnarray}
S_{I}^{D} &=&2q\left[ \overline{I}_{0}+\overline{I}_{L}\right] =2q\overline{I%
}\left( \frac{\overline{I}_{L}+\overline{I}_{0}}{\overline{I}_{L}-\overline{I%
}_{0}}\right) =  \nonumber \\
&&2q\overline{I}\coth \left( \frac{\Delta F_{n}}{2k_{B}T}\right) \text{,}
\label{SID}
\end{eqnarray}
where $\Delta F_{n}=F_{n}(L)-F_{n}(0)$ is the quasi-Fermi level drop across
the sample.

Since thermionic emission and diffusion are independent processes, one can
compute their joint contribution to the current fluctuations. We obtain 
\begin{eqnarray}
S_{I}^{TE,D} &=&\left( \frac{\overline{v}_{D}/v_{r}}{1+\overline{v}_{D}/v_{r}%
}\right) ^{2}S_{I}^{TE}+\left( \frac{1}{1+\overline{v}_{D}/v_{r}}\right)
^{2}S_{I}^{D}  \nonumber \\
&=&2q\left[ \overline{I}+2I_{s}\right] =2q\overline{I}\coth \left( \frac{q%
\overline{V}}{2k_{B}T}\right) \text{,}  \label{SITED}
\end{eqnarray}
which is of the full shot noise type. This result is just a manifestation of
the fact that both diffusion and thermionic emission processes are in origin
uncorrelated processes. Notice, that from the present derivation, this
result is valid for the whole range of applied bias.

Now we are in a position to develope a physical interpretation of the noise
properties of SBD, in particular of the cross-over between shot and thermal
noise behaviors. Let us start considering the case of low applied bias. As
is well known, in this range of bias the device is controlled by the
depletion layer present near the rectifying contact, where almost of the
applied potential drops. This implies that the quasi-Fermi level is nearly
constant all over the sample, and hence, according to Eq.(\ref{H(x)}), $H(x)$
is also (nearly) constant. Therefore, we may neglect the contribution of the
long range Coulomb interaction term $\delta I_{E}(t)$. Accordingly, the
current fluctuations are determined by the diffusion and thermionic emission
terms, thus giving rise to a spectrum of the shot noise type (see Eq.(\ref
{SITED})). On the contrary, for higher applied bias the depletion layer
disappears and the applied bias drops all over the sample. In this situation
the quasi-Fermi level is no longer constant, and the self-consistent term $%
\delta I_{E}(t)$ plays a relevant role. This term introduces correlations
between the otherwise uncorrelated thermionic emission and diffusion
processes, thus suppressing the shot noise. Therefore, the thermal noise
associated to series resistance\cite{Gomila98} may be interpreted as just
suppressed shot noise due to long-range Coulomb interaction.

The previous results can be understood physically by considering the density
profile corresponding to these devices. At low voltages these devices
display a density distribution with a deep depletion layer near the
interface followed by a quasi-neutral region that extends up to the back
ohmic contact. Since the device is controlled by the depletion layer, one
only needs to take into account this part in the discussion. But, in the
depletion layer the number density becomes very small. As a consequence,
screening effects are very small and long range Coulomb interaction can be
neglected, in agreement with the mathematical prove presented above. On the
other hand at higher voltages, the density distribution is nearly uniform
and equal to the doping density, thus providing appreciable screening
effects, which are the responsible of suppressing the shot noise.

In summary, we have developed a theoretical approach to interpret physically
the noise properties of Schottky diodes. We have shown that the occurrence
of shot noise at low bias has to be attributed to the fact that the density
distribution is such as to make negligible the effects of the long range
Coulomb interaction. As a result the originally uncorrelated diffusion and
thermionic emission processes remain uncorrelated, thus giving rise to a
shot noise spectrum. On the other hand, at higher applied bias the nearly
uniform carrier density profile induces considerable screening effects,
which introduce correlation between the diffusion and thermionic emission
process. The resulting thermal noise spectrum is thus the result of the
suppression of the intrinsic shot noise of the device. The main ideas
developed in the present paper can be extended to other devices, like $p-n$
diodes, tunneling diodes, ballistic devices or mesoscopic devices, thus
offering new perspectives on what concerns the understanding of the noise
properties of semiconductor devices.

\vspace{0.5cm}\noindent Acknowledgments\noindent

Drs. T. Gonz\'{a}lez and O. M. Bulashenko are thanked for stimulating
discussions on the subject. Partial support from the Italian MURST ''Physics
of Nanostructures'', from the Spanish SEUID and from the NATO linkage grant
HTECH.LG 974610, is gratefully acknowledged.

\end{multicols}


\begin{references}
\bibitem{remark}  Indeed, above flat band conditions the electric field
inside the sample can be assumed to be nearly constant, thus we can
approximate $\int_{0}^{x}\overline{E}(x^{\prime })dx^{\prime }\approx (%
\overline{V}-V_{bi})x/L$. By substituting this expression into Eq.(\ref
{vD(t)}), one obtains $\overline{v}_{D}=\mu (\overline{V}%
-V_{bi})N_{D}/(n^{eq}L)\exp (-q\beta \overline{V})$, where we have used the
expression for $V_{bi}$, Einstein's relation and the fact that $\exp (q\beta
(\overline{V}-V_{bi}))\gg 1$. This last condition also implies $\overline{v}%
_{D}\ll v_{r}$, thus giving $I_{s}\approx qA\mu N_{D}/L(\overline{V}%
-V_{bi})\exp (-q\beta \overline{V})$.
\end{references}
\end{document}